# Spatial variations in the Caspian Sea wave climate in 2002–2013 from satellite altimetry

Nadezhda Kudryavtseva[a], Kuanysh Kussembayeva[b], Zaure B. Rakisheva[b] and Tarmo Soomere[a,c]

[a] Department of Cybernetics, School of Science, Tallinn University of Technology, Akadeemia tee 21, 12618 Tallinn, Estonia; nadezhda.kudryavtseva@taltech.ee
[b] Al-Farabi Kazakh National University, Department of Mechanics, al-Farabi av. 71, 050038 Almaty, Kazakhstan
[c] Estonian Academy of Sciences, Kohtu 6, 10130 Tallinn, Estonia

**Abstract.** The core properties of the wave climate and its changes in the Caspian Sea are established in terms of the annual mean significant wave height and its regional changes in 2002–2013 based on the outcome of the satellite altimetry mission JASON-1. Remotely estimated wave heights are validated against properties of the empirical distribution of instrumentally measured wave heights in the southern Caspian Sea and monthly averages of visually observed wave heights at three locations. A correction for systematic differences leads to very good correspondence between monthly averaged *in situ* and satellite data with a typical root mean square difference of 0.06 m.

The average significant wave height in the Caspian Sea is 0.5–0.7 m in the northern basin of the sea, around 1.2 m in large parts of the central and southern basins and reaches up to 1.8 m in the northern segment of the central basin. The basin-wide average wave intensity varied insignificantly in the range of 1.02–1.14 m in 2002–2013. These estimates overestimate the wave heights by about 30% because low wave conditions are ignored. Substantial and statistically significant changes in the wave height occurred in certain areas. The wave height decreased by $0.019 \pm 0.007$ m/yr in the eastern segment of the central basin and by $0.04 \pm 0.04$ m/yr in the western segment of the southern basin. These changes can be explained by an increase in the frequency of westerly winds at the expence of southerly winds. Both basin-wide and regional extreme wave heights exhibit large interannual variations but do not show any significant trend. The patterns of changes in mean and extreme wave height are different. The average wave height has increased while the extreme wave height has decreased in the eastern segment of the southern basin.

## INTRODUCTION

The Caspian Sea (Kostianoy & Kosarev 2005) is the largest inland water body in the world with a surface area of 371 000 km$^2$ (Fig. 1). Its size is comparable to the surface of several shelf or semi-enclosed seas connected to the World Ocean such as the Baltic Sea, Black Sea or the Red Sea, and substantially exceeds the area of, e.g., the entire system of Great Lakes. As the Caspian Sea is a practically non-tidal water body, its shores are mostly developing under the joint impact of wind waves, long-term variations in the water volume of the entire sea and local storm-driven short-term changes in the water level.

The Caspian Sea has a more or less regular shape, with a maximum length in the north–south direction of about 1000 km and a width of about 600 km (Kostianoy & Kosarev 2005). It has experienced substantial changes in the water level over the past 120 years (Chen et al. 2017a, 2017b), predominantly caused by changes in the Volga River discharge. In 1900–1940 the sea level was steadily falling from –25.6 m in 1900 to –27.7 m in 1941 when the sea level stabilized. In 1950–1970 the Caspian Sea region experienced severe anthropogenic impact from the building of several hydroelectric power stations (HPSs) on the Volga, such as Nizhegorodskaya HPS (1956), Volga HPS (1961) and Zhigulevskaya HPS (1957). These structures caused a shift in the seasonal pattern of water level variations in the Caspian Sea. The peak in the Volga River discharge and, accordingly, in the Caspian Sea water level, shifted from June to May. This was accompanied by a further drop in the average water level down to –28.3 (Yaitskaya 2017). In total, eight HPSs were built on the Volga River between 1940 and 1986. The sea level continued to fall also in the 1970s and reached as low value as –29.0 m in 1977, but then increased in 1984–1996 and stabilized at the level of –27.1 m in recent years (1997–2015). Figure 2 shows a summary of sea-level changes together with an overview of changes in the hydrometeorological conditions in the Caspian Sea region.

The changes in the water level apparently had no distinguishable effects on the geometry and wave regime of the relatively deep-water central and southern basins of the sea where the depth reaches approximately 800 and 1000 m, respectively, and where the nearshore slope is relatively large (e.g., Yaitskaya 2017). However, recognizable changes may have occurred in the shallow northern basin of the sea that has extensive areas with water depth less than 25 m. Figure 1 (right) shows a map of the Caspian Sea bathymetry constructed using the ETOPO1 Global Relief Model data (Amante & Eakins 2009).

The information about the wave climate of the Caspian Sea is sparse and fragmented (e.g., Vignudelli et al. 2008). The most comprehensive information about the fundamental properties of the wave climate in this sea is presented in contemporary atlases of wind and waves for the last 30 years (Lopatoukhin et al. 2003; Boukhanovsky et al. 2011). The wave climate is moderate with the average wave height up to 1.2 m (e.g., Myslenkov et al. 2018) and wave properties exhibit strong seasonality in this water body (Rakisheva et al. 2019). Typically to sheltered and semi-sheltered water bodies of this size, the wave fields contain a relatively small contribution of the swell.

Such atlases, however, do not discuss temporal variations in the wave climate. They only provide the main parameters of the wave regime averaged over large spatial domains and thus have a poor spatial resolution. The results of the eleven years (1992–2002) of wave simulation in the southern Caspian Sea offered a better resolution in space but moderate coverage in time (Golshani et al. 2007). Numerical simulations using a model with an about 8 km resolution indicated the presence of considerably different wave regimes in each of the three main basins, namely the northern, central and southern basins. Recent, more detailed studies of Caspian Sea wave fields are essentially various attempts to evaluate the wave energy potential in different locations of the sea (Nejad et al. 2013; Zamani & Badri 2015; Kamranzad et al. 2016; Amirinia et al. 2017). Due to the above features of the wave climate several studies have come to a conclusion that wave energy is irrelevant in this water body compared to the wind energy (Rusu & Onea 2013). However, similarly to other semi-sheltered seas with complicated geometry (Kovaleva et al. 2017), the nearshore wave regime of the Caspian Sea contains hot spots of high wave energy concentration (Hadadpour et al. 2014).

Higher-resolution studies of the wave properties in this water body (Bruneau & Toumi 2016; Myslenkov et al. 2018) with a model resolution up to 1 km in the coastal zone and/or using the global hydrometeorological data reanalysis (Lopatoukhin & Yaitskaya 2019) have shed further light on the mean and extremes of wave properties. The simulations have demonstrated virtually no changes in the storm activity in the Caspian Sea region (Myslenkov et al. 2018). The maximum wave height with a probability of exceedance of 3% may reach 11.7 m (Myslenkov et al. 2018). Zounemat-Kermani & Kisi (2015) employed an alternative approach to understand the main features of the wave climate using chaos theory. They showed that all basic wave properties, such as height, period and direction, exhibit the presence of high-dimensional chaotic behaviour in the Caspian Sea.

The meteorological conditions have significantly changed in the Caspian Sea region over the last 65 years (Fig. 2). For example, in the northern and central parts of the sea, westerly winds were predominant in 1950–1970 with a characteristic wind speed of 2–4 m/s. A shift in wind properties occurred in 1971–1983. The gradual predomination of southerly and southeasterly winds was accompanied by a decrease in the wind speed. No change in the wind direction and speed was observed in 1984–1996. In 1997–2015 the wind direction shifted back to westerly and northwesterly winds as in 1971–1983. This process was accompanied by an increase in the wind speed, indicating a cyclic behaviour on a timescale of ~26 years. The southern part of the Caspian Sea also experienced a shift in the predominant wind direction from the southwest to the south in 1971–1983, and then back to the southwest in 1997–2015.

A shift in wind direction in small semi-sheltered water bodies with a complicated shape can have a drastic effect on the wave conditions due to changes in the fetch length. A shift in the wind direction in 1971–1983 (Fig. 2) caused a change in the wave conditions in the Caspian Sea with a possible increase in the wave heights in the central part and a decrease in the northern and southern parts (Yaitskaya 2017). However, no quantitative studies of the effect of shifts in wind direction on wave heights are available in the international literature so far.

Very few contemporary instrumental wave records are known for the Caspian Sea. The existing data are not widely accessible (Hartgerink 2005) and often only available for researchers within the country that holds the data. According to Ambrosimov (2008), such records seem only to exist starting from the year 2005. For this reason, the employment of an alternative source of wave data, such as satellite altimetry (e.g., Hemer et al. 2007), is essential in order to establish an adequate picture of the wave climate and its variations in this water body. The use of this source is complicated in relatively small

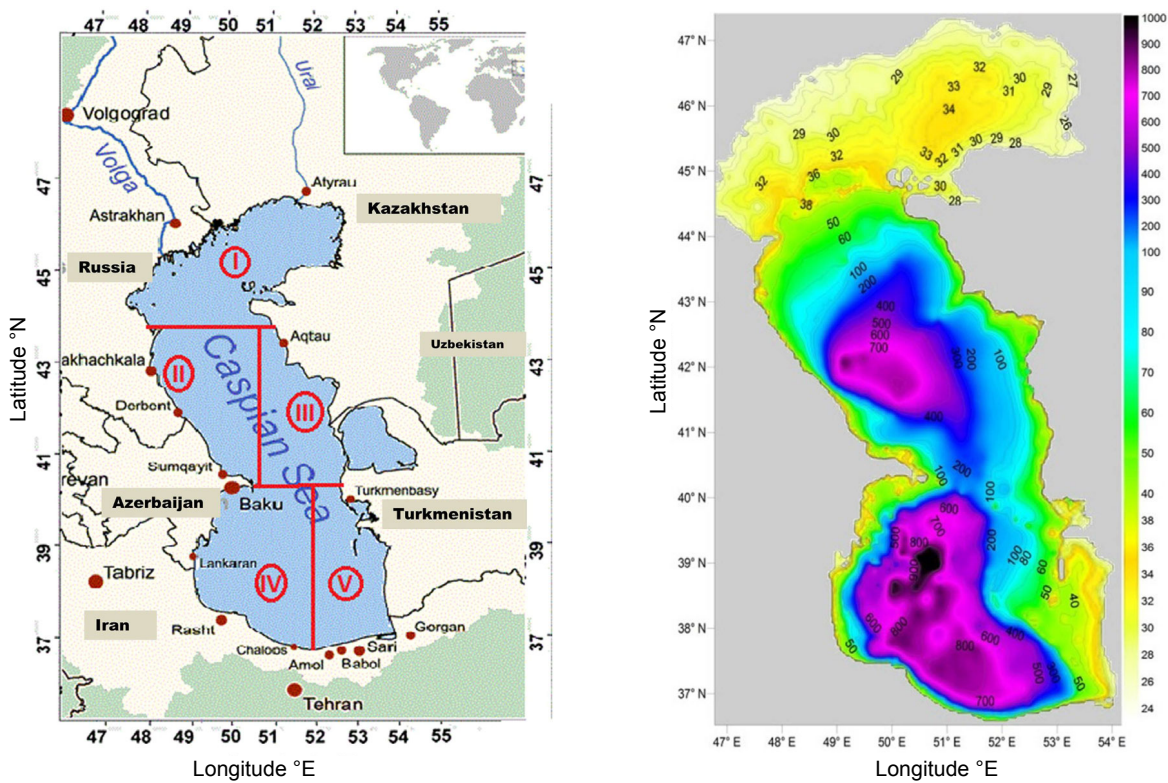

**Fig. 1.** The Caspian Sea is divided into five parts (left) with different wave climates and possibly different patterns of changes in wave properties (Rakisheva et al. 2019). The subregions are indicated by solid red lines: the northern part I (44°–47°N), the middle-western part II (40.5°–44°N, 47°–50.5°E), the middle-eastern part III (40.5°–44°N, 50.5°–53°E), the southwestern part IV (37°–40.5°N, 49°–52°E) and the southeastern part V (37°–40.5°N, 52°–54°E). As waves generated in the shallow Garabogazköl Basin practically do not reach the open Caspian Sea, this bay is excluded from consideration. The map of the bathymetry of the Caspian Sea (right); contours mark water depth in metres.

| Parameter | 1950–1970 | 1971–1983 | 1984–1996 | 1997–2015 |
|---|---|---|---|---|
| Water level | Decreasing ↓ | Decreasing ↓ | Increasing ↑ | Stable — |
| Wind direction | Northern and central parts: W (western) wind Southern part: SW | *Shift in wind direction* Northern and central parts: W changed to S and SE Southern part: SW changed to S | No change | *Shift in wind direction* Northern and central parts: S and SE changed to W and NW Southern part: S changed to SW |
| Wind speed | Stable — 2-4 m/s | Decreasing ↓ | Stable — ~2 m/s | Increasing after 2000 ↑ |
| Wave height | Stable? | Decreasing in the northern and southern parts ↓ Increasing in the central part ↑ | Stable — | Increasing? |

Building of HPSs on the Volga River | Shift in wind direction | | Shift in wind direction

**Fig. 2.** A summary of changes in hydrometeorological conditions in the Caspian Sea region in 1950–2015, following Yaitskaya (2017).

and seasonally ice-covered basins such as the Baltic Sea or the Caspian Sea. However, recent analysis has demonstrated the potential of this technique for estimates of several parameters of wave fields and wave climate also in basins similar to the Caspian Sea (Kudryavtseva & Soomere 2016, 2017).

Methods of satellite survey have been employed for the Caspian Sea for many purposes, including the tracking of oil pollution (Mityagina & Lavrova 2015), sea-level variability (Cazenave et al. 1997; Lebedev & Kostianoy 2008; Lebedev 2015), detection of internal waves (Lavrova et al. 2011; Mityagina & Lavrova 2016) or identification of various types of synoptic eddies (Karimova 2012). The complexity of the internal wave field signals that in many occasions their generation may be associated with the presence of intense surface waves.

In this paper we explore the possibilities of the use of satellite altimetry to derive reliable information about regional changes in the wave climate of the Caspian Sea and to establish whether the wave heights have followed certain trends in recent years. The aim is to (i) systematically evaluate the core parameters of the Caspian Sea wave climate, such as the spatial pattern of the annual average wave height in the entire basin and (ii) identify possible reflections of climate change in terms of this parameter, both basin-wide effects and changes in specific subbasins. The analysis is based on the outcome of satellite altimetry, specifically, on the estimates of the local significant wave height in 2002–2013 by the JASON-1 mission. The remotely sensed data are validated against the results of a selection of existing buoy measurements and visual observations. Since the period of observations in this study is eleven years and thus rather short for studying long-term changes in the wave climate, we use the notion 'wave climate' to denote also short-term wave climate variability.

## DATA AND METHODS

We use the satellite altimetry *Ku* band wave height data from the Radar Altimeter Database System (RADS) database (http://rads.tudelft.nl/rads/rads.shtml) (Scharroo et al. 2013). For this paper, the wave height data set from the JASON-1 satellite (Ménard et al. 2003) was selected. JASON-1 is also widely used as a reference satellite for multiple missions because it exhibits the minimal temporal drift in the measured parameters (e.g., Dibarboure et al. 2011).

The local measurements on the Caspian Sea were selected from the JASON-1 full data set using the boundaries of 46°–55°N for longitudes and 36°–48°E for latitudes. The data are from 2002 to 2013, spanning more than ten years. JASON-1 has very good spatial coverage over the Caspian Sea (Fig. 3). Although the altimeter tracks do not cover the Caspian Sea entirely during every single month (Fig. 3, left panel), the excellent long-term coverage makes the outcome of the JASON-1 mission a unique data set for studies of the spatial distribution of the Caspian Sea wave properties and spatial changes in the wave climate.

The size, overall shape and to some extent also the complexity of the shores of the Caspian Sea are similar to those of the Baltic Sea. Both seas are also partially ice-covered (Leppäranta & Myrberg 2009). Therefore, it is natural to assume that the limitations for the quality of satellite-derived wave information from the RADS database established for the Baltic Sea conditions (Kudryavtseva & Soomere 2016) are applicable also in the Caspian Sea conditions. This study is based on the JASON-1 data that showed the best performance in the Baltic Sea when compared with *in situ* measurements (Kudryavtseva & Soomere 2016).

Previous research in the Baltic Sea region has shown that properties of waves extracted from satellite altimetry for the snapshots that are located at a distance <0.2° from the shore (including shores of any of the islands) are generally erroneous (Kudryavtseva & Soomere 2016). Such data entries (shown in red in Fig. 3, right panel) were excluded from the analysis. For this purpose, we used the shoreline location from the GSHHG database (Global Self-consistent Hierarchical High-resolution Geography, version 2.3.6 August 19, 2016, https://www.ngdc.noaa.gov/mgg/shorelines/, last accessed 04.09.2019) provided by the National Oceanic and Atmospheric Administration (NOAA), National Centres for Environmental Information (NCEI), National Oceanic and Atmospheric Administration, U.S. Department of Commerce. The details of the processing and assembly of the GSHHG shorelines are described in Wessel & Smith (1996).

The data entries with the values of the backscatter coefficient >13.5 were also excluded, following the limitations found for the Baltic Sea (Kudryavtseva & Soomere 2016). The use of this threshold is equivalent to cutting off calm conditions with wind speeds below 2.5 m/s. This procedure removes situations with low waves from the analysis and thus introduces a certain bias to some characteristics of the wave climate, such as the average wave height, and may also distort to some extent the seasonal, interannual or spatial variability in the wave climate. However, other core properties such as higher quantiles and extreme wave heights, qualitative patterns of spatio-temporal variations and trends in wave heights are not significantly affected.

We also removed from consideration the data points that contained large errors of >0.5 m in the normalized standard deviation in the significant wave height. As discussed in Kudryavtseva & Soomere (2016), the ice flagging at 50% ice concentration (based on EUMETSAT OSI SAF Global Sea Ice Concentration Reprocessing data OSI-409-a data set) is highly consistent for the Baltic Sea in the RADS database. It is, therefore, natural to assume that the satellite data from the Caspian Sea region also have consistent information about the ice conditions. Based on this conjecture, we used the standard ice flagging provided by the RADS data set to exclude situations with extensive ice cover.

Taking into account limited capabilities of validation with the *in situ* data in the Caspian Sea region, we also extensively checked the single data entries for the erroneous measurements and outliers. We introduced an additional data quality check based on the difference between the significant wave heights $H_S$ evaluated using $C$ and $Ku$ bands. Extensive tests showed that excluding the measurements with the difference $|H_{SC} - H_{SKu}| > 1.7$ m removes most of the erroneous entries.

**VALIDATION**

**Overview of *in situ* wave height data in the Caspian Sea**

In order to reach substantiated conclusions about wave climate and its changes, it is essential to verify the quality of the outcome of satellite altimetry against ground truth in the target area. Such a comparison is complicated in the Caspian Sea because of a scarcity of *in situ* wave observations and measurements and limited availability of the relevant data. The search of such data through contacts with the organizations responsible for coastal and marine observations in the Caspian Sea countries showed that there are at least 35 wave observation or measurement sites in the Caspian Sea (Fig. 4). Almost 2/3 of the sites (22 out of 35, or 63%) perform only visual wave observations.

Table 1 lists the location, World Meteorological Organization (WMO) code, geographical coordinates of the stations together with the organization which owns the data, temporal coverage, type and frequency of observations. Recently, monthly averaged wave heights of 13 stations with the observations in 1977–1991 and three stations with the measurements in 1977–2018 have become available at the Hydrometeorological Research Center of the Russian Federation website (www.esimo.ru). This data set, based on visual observations, is used in the paper for validation of the satellite data. It is expected that there will be a difference between the satellite-derived data and visual observations since the relevant estimates of wave height are performed in an entirely different way.

In Table 1, only stations operated by the Hydrometeorological Research Center of the Russian Federation have publicly available wave data. The information for these stations only reflects the period of observations in terms of available monthly data. At least some of these stations have likely been active during a much longer time interval similar to the coastal hydrometeorological stations in the Baltic countries (Soomere 2013). The information about the Iranian stations is very limited and does not cover the years of observations performed or how the wave properties are established. Therefore, only the names and coordinates of the Iranian stations are listed.

**Empirical distribution of instrumentally measured wave heights**

A direct comparison of instantaneous wave properties with the satellite altimetry based estimates is usually not straightforward, especially in the coastal regions. The footprint of the altimetry signal covers a relatively large offshore area and can have a time mismatch with the observation time of *in situ* stations. Besides, lack of open access individual wave height measurements makes it problematic to validate the satellite altimetry data in the Caspian Sea region.

The primary criterion for the robustness of the outcome of observations or measurements is that the wave properties are following a correct probability distribution. For this reason, we start from a comparison of the outcome of a series of measurements of wave heights in terms of empirical probability distributions of observed or measured wave heights. While single wave heights, ideally, follow a Rayleigh distribution (Longuet-Higgins 1952) that is possibly modified by nonlinear effects (Prevosto et al. 2000), especially in the nearshore (Cherneva et al. 2005), the snapshots of significant wave heights usually follow a (modified) Weibull distribution (Muraleedharan et al. 2007).

For validation of the appearance and properties of the empirical distribution of significant wave heights, we use published parameters of the wave regime at the Iranian station Anzali (Fig. 4). The wave heights were evaluated once an hour using the standard 20-min continuous measurements with an Oceanor discus buoy during a 13-month long time interval (09 May 2007–18 June 2008). The significant wave heights follow the Weibull distribution with the scale parameter $\alpha = 0.95$ and the shape parameter $\beta = 1.42$ (Zamani & Badri 2015). The Weibull distribution matched the empirical distribution more consistently than normal, lognormal or Rayleigh distribution at this site (Zamani & Badri 2015).

The JASON-1 satellite data were selected for the same period as the published data and within <0.5°

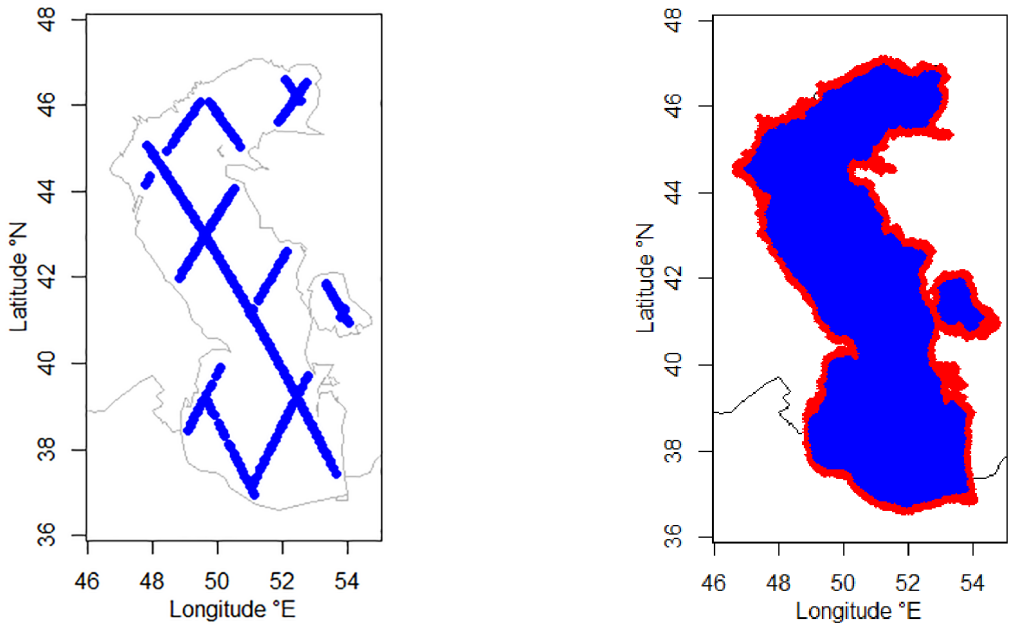

**Fig. 3.** JASON-1 tracks (blue lines) over the Caspian Sea during one month of observations (April 2008, left) and the whole period of 2002–2013 (right). The altimeter tracks are shown in blue. Red points show the measurements located within <0.2° from the coast which were excluded from the analysis.

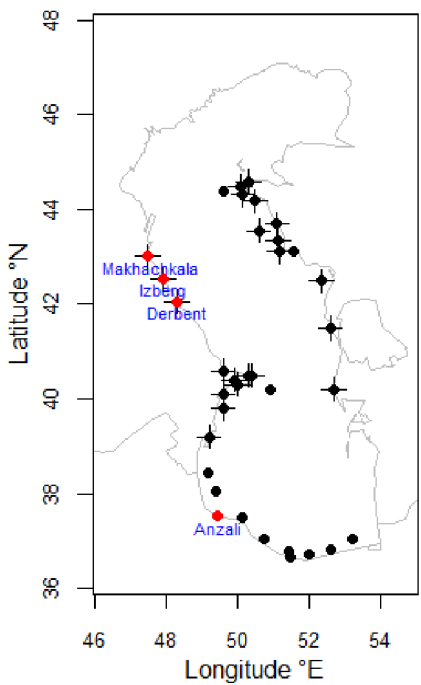

**Fig. 4.** Wave observation or measurement sites in the Caspian Sea (Table 1). The sites used for validation of JASON-1 wave heights are marked with red points and names in blue. Crosses indicate the measurement sites that perform only visual observation of wave heights.

distance from the Anzali station. The parameters of the Weibull distribution of the set of estimates of significant wave height from satellite altimetry showed a reasonably good fit to wave buoy data. The scale parameter of the satellite-derived distribution is $\alpha = 1.12 \pm 0.17$ and the shape parameter is $\beta = 1.75 \pm 0.33$. The estimates of these parameters from the buoy data (Zamani & Badri 2015) are thus within standard 95% error intervals of the satellite-based estimates. Variations in the threshold distance of the satellite snapshot from the coast did not significantly affect the results. It is, therefore, safe to say that the JASON-1 wave height data adequately represent both the appearance and basic parameters of the empirical distribution of significant wave heights in the nearshore of the Caspian Sea.

**Validation based on monthly averages of visually observed wave heights**

The satellite altimetry wave height data are further validated by comparing them with the available visual observations (Fig. 4). The JASON-1 wave heights represent the significant wave height $H_S$, which corresponds approximately to the crest-to-trough height of the 1/3 largest waves in the footprint. The effective footprint depends on $H_S$ (Chelton et al. 1989). For JASON-1, assuming the satellite altitude of 1335 km,

**Table 1.** Wave height measurement locations and the basic properties of the data sets in the Caspian Sea region. The code in brackets denotes the WMO numeric station code assigned to synoptic weather stations. The first number 9 of the code specifies the Pacific region. Monthly average wave heights for the stations operated by the Hydrometeorological Research Center of the Russian Federation (www.esimo.ru) are publicly available. There is no public access to other data sets. Stations used for the validation of the JASON-1 data are marked in bold. The information for Fort Shevchenko/Fort-Shevchenko (97060) and Shevchenko City/Aktau (97061) is presented according to national services of Russian Federation and Kazakhstan

| Station name, (code), country | Coordinates °N | Coordinates °E | Dates of observation | Type | Frequency |
|---|---|---|---|---|---|
| Hydrometeorological Research Center of the Russian Federation www.esimo.ru | | | | | |
| Fort Shevchenko (97060), Kazakhstan | 44.6 | 50.3 | 1977–1991 | Visual | |
| Shevchenko City, Aktau (97061), Kazakhstan | 43.7 | 51.1 | 1977–1991 | Visual | |
| Bektash (97065), Turkmenistan | 41.5 | 52.6 | 1977–1991 | Visual | |
| Kuuli-Mayak (97073), Turkmenistan | 40.2 | 52.7 | 1977–1991 | Visual | |
| Kultuk (97004), Azerbaijan | 39.2 | 49.2 | 1977–1991 | Visual | |
| Sengi-Mugan (97007), Azerbaijan | 39.8 | 49.6 | 1977–1991 | Visual | |
| Bulla (97009), Azerbaijan | 40.1 | 49.6 | 1977–1991 | Visual | |
| Neftyanye Kamni (97017), Azerbaijan | 40.2 | 50.9 | 1977–1991 | Buoy | Every 3 h |
| Peschannyi (97014), Azerbaijan | 40.3 | 50.0 | 1977–1991 | Visual | |
| Absheron (97093), Azerbaijan | 40.5 | 50.4 | 1977–1991 | Visual | |
| Darwin Bank (97019), Azerbaijan | 40.5 | 50.3 | 1977–1991 | Visual | |
| Baku (97012), Azerbaijan | 40.4 | 49.9 | 1977–1991 | Visual | |
| Sumgait (97021), Azerbaijan | 40.6 | 49.6 | 1977–1991 | Visual | |
| **Derbent (97025), Dagestan, Russia** | **42.06** | **48.3** | **1977–2018** | **Visual** | |
| **Izberg (97026), Dagestan, Russia** | **42.53** | **47.91** | **1977–2018** | **Visual** | |
| **Makhachkala (97027), Dagestan, Russia** | **43.016** | **47.48** | **1977–2018** | **Visual** | |
| 'Kazhydromet' Ministry of the Energy of the Republic of Kazakhstan www.kazhydromet.kz | | | | | |
| Kulaly (97059), Kazakhstan | 44.50 | 50.07 | Since 1936 | Visual | |
| Kuryk (97065), Kazakhstan | 43.56 | 50.59 | Since 2012 | Visual | |
| Saura (97064), Kazakhstan | 44.19 | 50.48 | Since 2008 | Visual | |
| Fetisovo (97063), Kazakhstan | 42.49 | 52.35 | Since 2002 | Visual | Every 6 h |
| Cape Peschany (97062), Kazakhstan | 43.11 | 51.16 | Since 2008 | Visual | |
| Aktau (97061), Kazakhstan | 43.36 | 51.13 | Since 1960 | Visual | |
| Fort-Shevchenko (97060), Kazakhstan | 44.32 | 50.14 | Since 1848 | Visual | |
| Kuryk, Kazakhstan | 43.12 | 51.57 | May–July 2018 | Buoy | Every hour |
| Buoy 2, Kazakhstan | 44.38 | 49.59 | Aug–Sept 2016 | Buoy | |
| Port and Maritime Organization (PMO), https://www.pmo.ir Iranian National Institute for Oceanography and Atmospheric Science (www.http://www.inio.ac.ir) | | | | | |
| Amirabad, Iran | 37.04 | 53.21 | – | – | |
| Babolsar, Iran | 36.82 | 52.625 | – | – | |
| Noor, Iran | 36.725 | 52.00 | – | – | |
| Chalus, Iran | 36.78 | 51.42 | – | – | |
| Ramsar, Iran | 37.03 | 50.75 | – | – | Every 6 h |
| Kiashahr, Iran | 37.49 | 50.13 | – | – | |
| **Anzali, Iran** | **37.52** | **49.45** | – | – | |
| Hashtpar, Iran | 38.06 | 49.40 | – | – | |
| Astara, Iran | 38.43 | 49.19 | – | – | |
| Bahonar Port and Maritime Administration, https://noshahrport.pmo.ir | | | | | |
| Noshahr, Iran | 36.65 | 51.50 | – | – | – |

– not available.

the effective width of the footprint in the Caspian Sea varies from 2 km ($H_S < 1$ m) to 7 km ($H_S \sim 5$ m). Therefore, depending on the wave height at the time of the measurement, the wave properties are averaged over different areas. This procedure often introduces a bias in the wave height, especially near the coast, as the wave height usually decreases towards the shore.

Visual observations are typically performed by observers who stand at the coast. The nearshore is usually shallow, and the presence of the seabed slope and various bathymetric features can create substantial modifications in the wave properties owing to wave refraction, reflection, shoaling, damping, and possibly sheltering for waves from specific directions (see Soomere 2013 and references therein). The quality of visual observations performed offshore, especially from the ships, has been thoroughly analysed (Soares 1986; Gulev & Hasse 1998, 1999). As the experienced observer's estimate of the wave height usually matches the significant wave height, the largest systematic deviations between the two data sets likely stem from the joint impact of the change in the footprint size with the wave height and modifications of the wave properties in the nearshore.

To validate the satellite altimetry data from the JASON-1 mission, we selected three stations with the period of observations matching the JASON-1 operating time. These measurement sites, namely Derbent (97025), Izberg (97026) and Makhachkala (97027), are located in Russia. Visual observations are performed at these stations every 3 h. For each station, we selected satellite altimetry data within a distance of 0.5°. From these data, we evaluated the monthly average significant wave heights. The average $H_{S,\text{sat}}$ was calculated as

$$H_{S,\text{sat}} = \frac{1}{N_m}\sum_{i=1}^{N_m} H_{S,\text{sat},i}, \qquad (1)$$

where $N_m$ is the number of satellite observations per month $m$ and $H_{S,\text{sat},i}$ denotes a single satellite altimetry derived estimate of wave height. To eliminate un-representative data clusters, the months with less than eight satellite measurements were excluded from the analysis. This cutoff value and the threshold for the minimum distance of the satellite snapshot from the station were thoroughly tested.

A direct comparison of the observed *in situ* and satellite-derived wave heights reveals a significant discrepancy. The root mean square difference (RMSD) between the two ranged from 0.8 m at Izberg and Makhachkala to 1.1 m at the Derbent measurement site (Fig. 5, left). Such a high RMSD between the outcome of visual observations and satellite altimetry data was also discussed in Lebedev & Kostianoy (2008).

Interestingly, there is a robust and systematic relationship between the difference of *in situ* observed and satellite-based wave heights, and the instantaneous severity of wave conditions. This is exemplified in the right panel of Fig. 5 at the example of Derbent wave data. The difference between monthly averages of satellite-derived and *in situ* visually observed wave height almost perfectly linearly depends on the background satellite-derived wave height. All three stations show the same systematic dependence and similar properties of the linear regression line. The parameters for such a linear fit for Derbent and Izberg practically coincide, whereas they are only slightly different for the Makhachkala measurement site (Table 2).

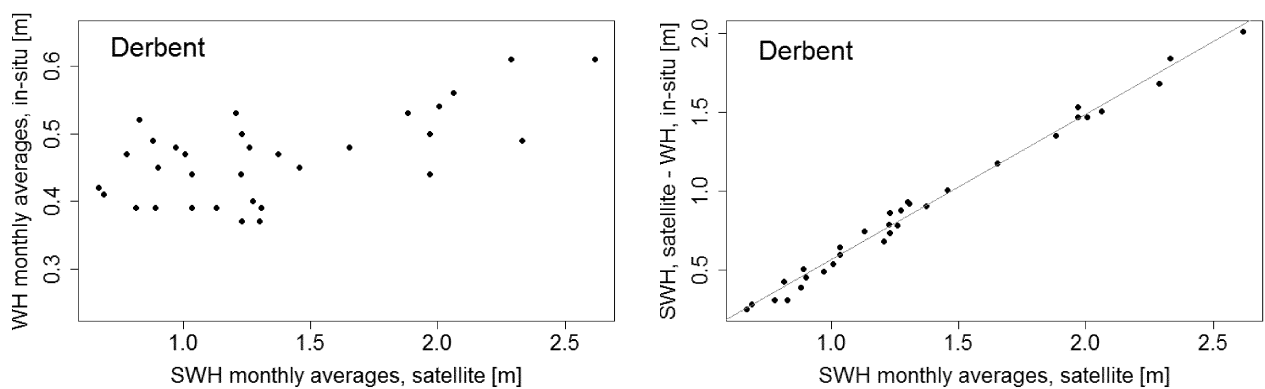

**Fig. 5.** Comparison of *in situ* and JASON-1 monthly averaged significant wave heights (SWH) for the Derbent site (left) and the difference between satellite-derived and *in situ* wave heights (WH) $H_{S,\text{sat}} - H_{in\ situ}$ and $H_{S,\text{sat}}$ from satellites (right). The solid line shows a fitted regression line (Table 2).

**Table 2.** Parameters of the linear regression line between monthly averaged values of $H_{S,\text{sat}} - H_{in\,situ}$ and $H_{S,\text{sat}}$ (Fig. 5, right) for the Derbent, Izberg and Makhachkala observation sites

| Station | Intercept $a$ | Slope $b$ | $p$-Value |
|---|---|---|---|
| Derbent | −0.36 ± 0.03 | 0.92 ± 0.02 | <$2.2e^{-16}$ |
| Izberg | −0.36 ± 0.03 | 0.90 ± 0.03 | <$2.2e^{-16}$ |
| Makhachkala | −0.24 ± 0.04 | 0.85 ± 0.03 | <$2.2e^{-16}$ |

Based on the established systematic, almost linear dependence, we adjusted the wave heights derived from satellite altimetry using the following linear expression:

$$H_{S,\text{sat}}^{*} = (1-a)H_{S,\text{sat}} - b, \qquad (2)$$

where the coefficients $a$ and $b$ are from Table 2. In essence, this adjustment mirrors the decrease in the wave height from the offshore to the observation site in the nearshore. Not surprisingly, the adjusted values $H_{S,\text{sat}}^{*}$ and monthly means derived from *in situ* observations after the correction show excellent correspondence. The resulting RMSD is 0.06 m for all three stations. Its single values follow a Gaussian distribution of differences between *in situ* observed and satellite-derived wave heights (Fig. 6). This feature indicates that, most likely, the primary source of errors in the adjusted data is random noise caused by the measurement uncertainty. Very small $p$-values (Table 2) emphasize the 'strength' of the correspondence in question.

### RESULTS

The local average significant wave height $H_S$, based on the data from the JASON-1 mission, is in the range of 0.4–1.8 m in the Caspian Sea in 2002–2013 (Fig. 7, left). The severest wave conditions occur, on average, in the northern segment of region II, next to Derbent, where the average wave heights reach about 1.8 m. Regions III and IV in the middle-eastern part and the southwestern part of the sea exhibit somewhat less severe average wave conditions with $H_S \sim 1.2$ m. Much lower average wave heights of about 0.5–0.7 m are found in region I in the northern basin of this water body. Such a low intensity of waves in this area probably reflects the very shallow depth of the region. A map of the number of satellite observations per grid cell is shown in Fig. 7 (right). Most of the cells have a similar number of measurements with a median of 301 observations per pixel, except for a few lines going from southeast to northwest and from southwest to northeast, which are characterized by up to 3000 measurements per pixel.

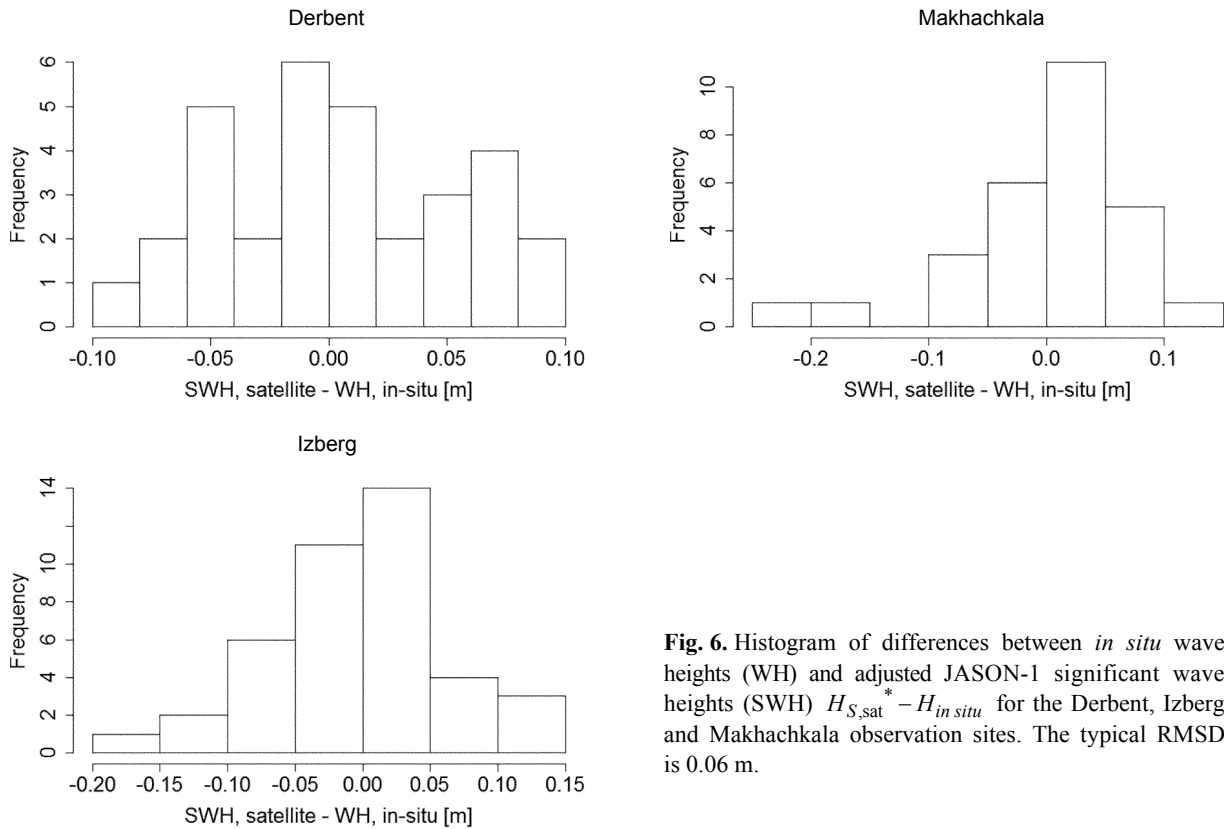

**Fig. 6.** Histogram of differences between *in situ* wave heights (WH) and adjusted JASON-1 significant wave heights (SWH) $H_{S,\text{sat}}^{*} - H_{in\,situ}$ for the Derbent, Izberg and Makhachkala observation sites. The typical RMSD is 0.06 m.

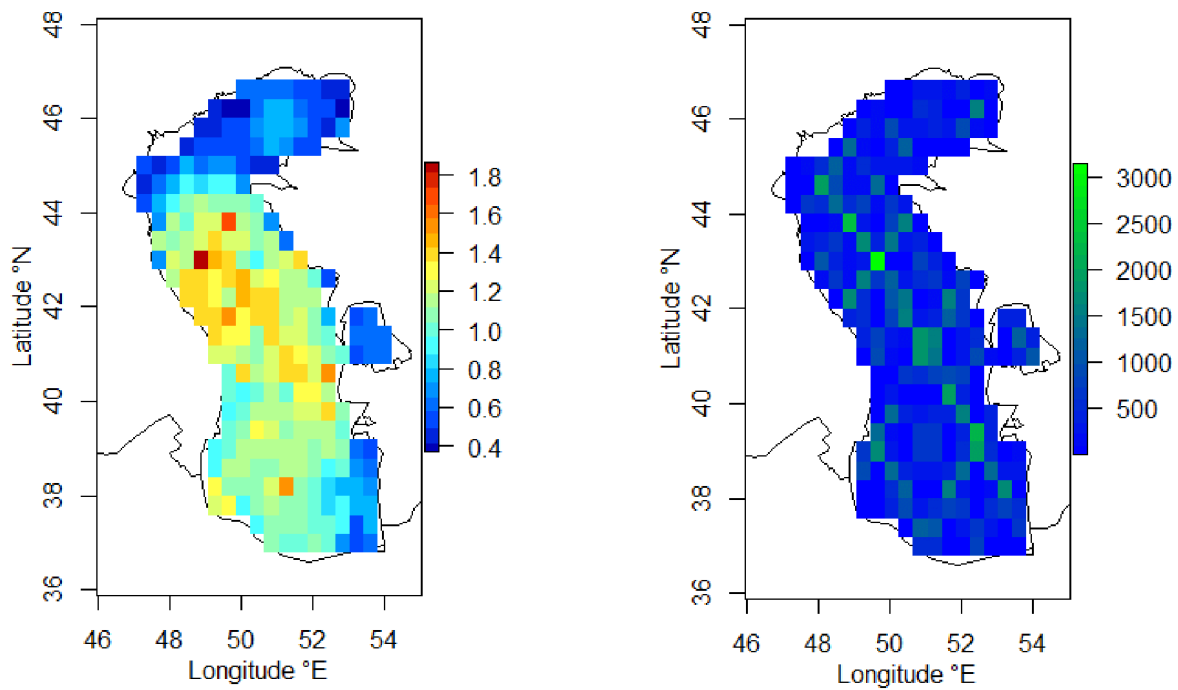

**Fig. 7.** Average significant wave height $H_{S,\text{sat}}$ (m, colour scale) in the Caspian Sea in 2002–2013 from JASON-1 satellite altimetry (left), and the number of satellite measurements per grid cell (right). The pixel size is 0.4° × 0.4° (18 × 25 pixels grid).

The presented estimates of average wave heights are most likely overestimated because we have removed a large part of relatively calm sea conditions from the analysis. However, the resulting map is expected to reflect well the spatial variations in the wave climate (Kudryavtseva & Soomere 2017). The obtained spatial structure of the Caspian Sea wave climate is consistent with the previous studies (e.g., Boukhanovsky et al. 2011; Lopatoukhin et al. 2003).

For each year, JASON-1 performed an almost equal number of measurements of the Caspian Sea. The resulting data sets contain on average 11 100 measurements per year except for the last years 2012 and 2013 when fewer measurements were performed (Fig. 8, left). The number of observations is by 25% smaller than the average in 2012 and by 62% in 2013. This variation in the number of measurements can introduce a bias in the yearly averaged significant wave height and estimates of its linear trend. A test was performed to check how the exclusion of the last year can affect the results. The changes were fairly small and did not affect the conclusions of the paper.

The yearly average satellite-derived significant wave height for the entire Caspian Sea (covering all regions) is in the range of 1.02–1.14 m (Fig. 8, right). This quantity exhibits variations with a magnitude of about 0.1 m and on timescales of about three years. The weak decreasing linear trend of the annual mean wave height is not statistically significant ($p \sim 0.6$) in the sense that the probability of its slope to be nonzero is fairly small (about 60%). The confidence intervals shown in Fig. 8 characterize the uncertainties of annual estimates. They are calculated in a standard manner as errors of the mean value. The uncertainties are typically about ±0.01 m.

More detailed analysis reveals that the wave climate in some segments of the Caspian Sea has significantly changed and that these changes have a strongly varying spatial pattern. To assess such changes, we consider the satellite-derived significant wave heights separately for the five regions of this water body (Fig. 1, left), following Rakisheva et al. (2019) and the perception that in some similar seas such as the Baltic Sea a large part of changes in the wave properties are caused by changes in the directional structure of winds (e.g., Kudryavtseva & Soomere 2017). This partitioning includes a division of the central and the southern basin of the sea into two subbasins (II and III, and IV and V, respectively). This approach makes it possible to detect the potential impact of changes in both the meridional and zonal components of winds on the properties of wind waves. This division results in a similar number of satellite observations per region, from 22 000 to 29 000 for regions I, II, III and V. Region IV has about 15 000 measurements. The number of observations per year varies from 1500 to 3000 and is rather stable between years, within 13% of variations for all regions except for region IV which exhibited more considerable variations (23%) in the number of satellite observations.

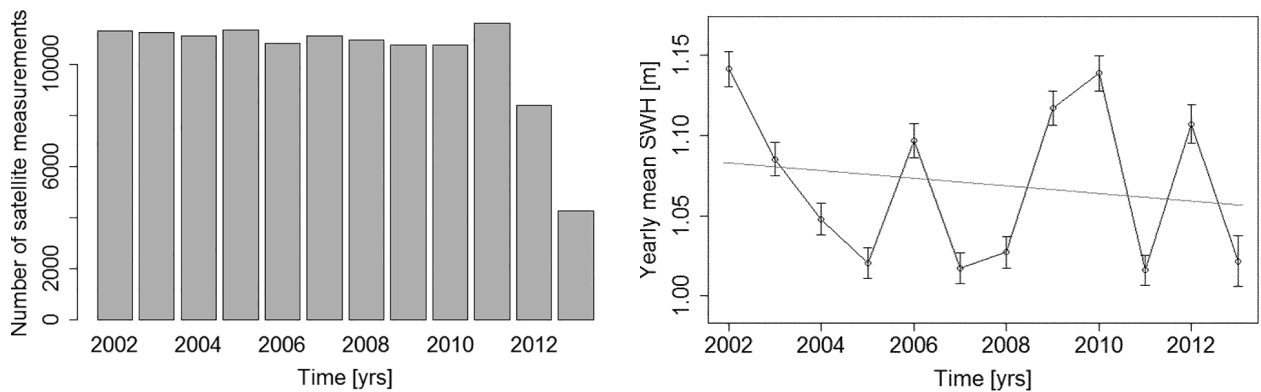

**Fig. 8.** The number of JASON-1 observations used for each year (left) and the annual mean significant wave height (SWH) in the entire Caspian Sea (right). The solid line shows the linear trend.

Interestingly, the yearly averaged significant wave height exhibits very different behaviour in different parts of the sea. This behaviour includes even the presence of different signs of the relevant trends in different regions (Fig. 9). The annual average significant wave height has slightly increased in regions I and IV that are located at different ends of the Caspian Sea. These trends are statistically not significant but indicate a certain increase in the frequency and/or intensity of meridionally aligned winds. A decreasing trend of comparable magnitude is formally present in region III but is also not statistically significant. The changes in the wave height in this region seem to contain extensive intra-decadal variations.

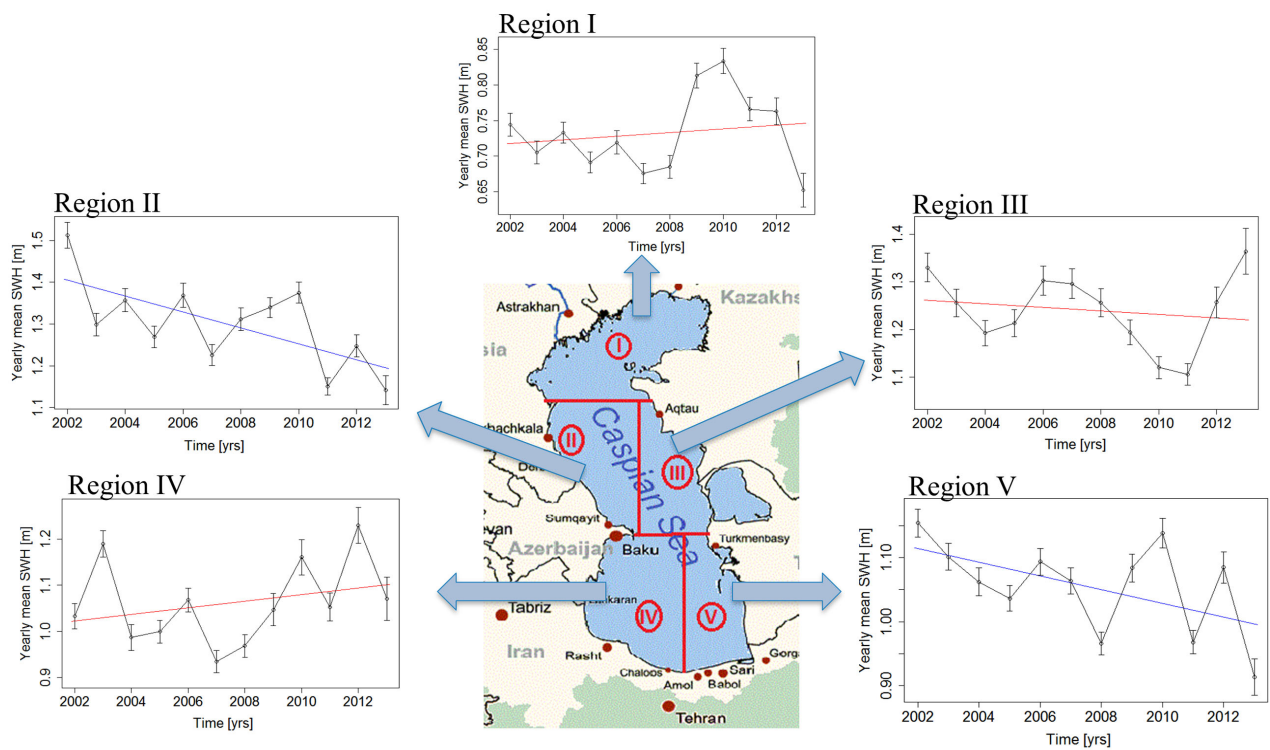

**Fig. 9.** Variations in the annual mean significant wave height (SWH) for five regions of the Caspian Sea. The solid red lines show linear trends (regression lines) that are statistically not significant. The solid blue lines show trends that are nonzero at a >95% level of statistical significance. The shown uncertainty is inversely proportional to the number of measurements per year: the lower is the number of measurement, the higher is uncertainty.

The annual average significant wave height has considerably decreased in regions II and V. These regions are located at differently oriented shores of the sea. Region II covers the western part of the central basin and region V is located in the eastern segment of the southern basin of the sea. The decrease in both regions II and V is significant in the sense that the slopes of trend lines are nonzero at a high level of statistical significance. The slope of this trend in region II is –0.019 ± 0.007 m/yr and is nonzero at a 99% level of statistical significance. This slope for region V, –0.04 ± 0.04 m/yr, is steeper but is nonzero at a 95% level of statistical significance because of a broader level of interannual variations and, therefore, larger uncertainty in the wave height in this region.

Differently from the Baltic Sea where the changes in the average and extreme wave heights have the same spatial pattern (Soomere & Räämet 2014; Soomere 2016), some changes in the severest wave conditions in the Caspian Sea are inconsistent with similar changes in the mean wave properties. The above has shown that the mean wave heights increased in region IV, but the 99th percentile shows an apparent decrease with time (Fig. 10). A decrease in the 99th percentile of the wave heights is characteristic of the years under question for most of the Caspian Sea except for region I. As the typical relative interannual variation in this percentile (up to 1.5 m or up to 40% of the long-term average) is much larger than the similar variation for annual mean wave heights, it is not unexpected that none of the discussed trends in the yearly averaged 99th percentiles is statistically significant.

**DISCUSSION AND CONCLUSIONS**

We have established the core properties of the wave climate and its changes in the Caspian Sea in terms of annual mean significant wave heights and regional patterns of changes of this quantity. The analysis is based on the estimates of significant wave height from the longest satellite altimetry mission JASON-1 from 2002 to 2013. This data set spans over more than ten years and has proved to be highly reliable in seasonally ice-covered inland seas with complicated geometry (Kudryavtseva & Soomere 2016).

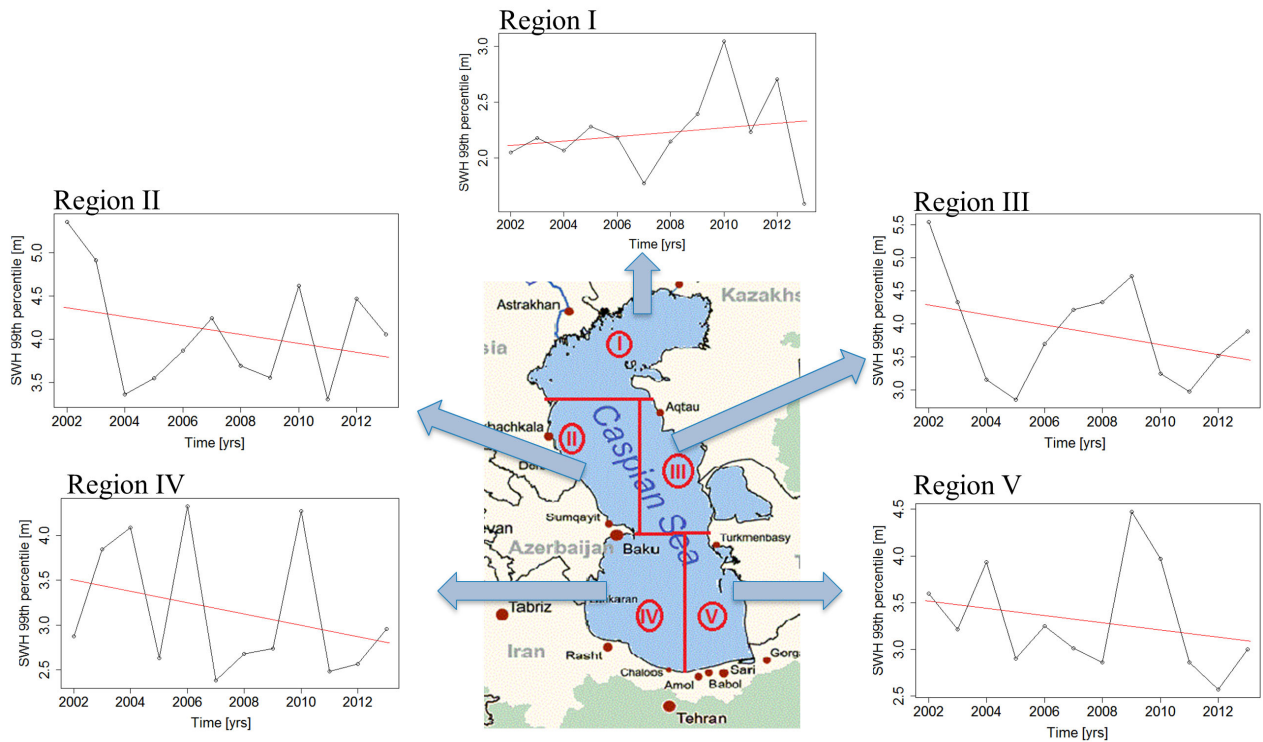

**Fig. 10.** Variations in the annual extreme wave heights (defined as the 99th percentile) of significant wave height (SWH) in five regions of the Caspian Sea. None of the linear trends (solid red lines) is statistically significant.

The estimates of wave heights derived from the JASON-1 mission are first validated against properties of the empirical distribution of significant wave heights extracted from buoy measurements in the southern nearshore of the Caspian Sea. The parameters of the empirical Weibull distribution fall into the standard error intervals of the similar distribution derived from satellite altimetry.

A further comparison of monthly average satellite-derived wave heights with similar visually observed wave properties at Derbent, Izberg and Makhachkala observation sites reveals an almost perfect linear relationship between the difference of these estimates and the background offshore wave height. The relevant adjustment of the offshore wave height results in an excellent correspondence between visually observed and satellite-derived wave properties.

Based on these two exercises, it is safe to say that satellite-derived wave heights of the JASON-1 mission reflect well the actual properties of the Caspian Sea wave fields. As low wave conditions are not recognized by satellite altimetry, the estimates based on the outcome of the JASON-1 mission apparently overestimate by ~30% the average wave heights but adequately reflect spatial patterns of wave intensity and catch the basic features of changes to mean and extreme wave heights. For example, our analysis detects the maximum average wave heights in the region centred at 50°E, 43°N, with average wave heights of ~1.6 m (Fig. 7). This location coincides with the one identified in Myslenkov et al. (2018) at 50°E, 42.8°N. The modelled average wave height in this location is ~1.2 m, which is 30% lower than obtained from satellite altimetry. The reported average wave height in Makhachkala is 0.7 m in 1997–2015 (Yaitskaya 2017), whereas the satellite altimetry data indicate average wave heights of ~1.2 m in the vicinity of Makhachkala.

The satellite-derived average significant wave height exhibits large spatial variability in the Caspian Sea. It is the largest (up to 1.8 m) in the northern segment of the central basin next to Derbent. Other segments of the central and southern basins have less severe wave conditions with the average significant wave height around 1.2 m. The average wave heights are much lower, typically about 0.5–0.7 m, in the northern basin of this water body.

For the first time, the trends in average wave heights in different regions of the Caspian Sea were studied. The average wave intensity in the entire basin varied in the range of 1.02–1.14 m and did not show any significant trend, but substantial changes have occurred in certain areas. The wave height has decreased by 0.019 m/yr in the eastern segment of the central basin and by 0.04 m/yr in the western segment of the southern basin. The relevant trends are nonzero at a 95% and 99% level of statistical significance, respectively. The average wave heights have slightly increased in the northernmost and southernmost ends of the sea. This spatial pattern may, hypothetically, be caused by an increase in the frequency and/or strength of the meridional wind component on the expense of easterly winds in the northern part of the sea and on the expense of westerly winds in the southern part of the sea.

The observed decrease in wave heights in regions II and III is consistent with the change in the predominant wind direction from southerly to southeasterly winds to westerly and northwesterly ones (Fig. 2). Southerly winds are characterized by the maximum fetch length of up to ~800 km (estimated from Chalus, Iran, to Aktau, Kazakhstan). The described shift in the wind direction can drastically reduce the maximum fetch length, down to ~300 km (from Makhachkala, Russia, to Aktau, Kazakhstan), which can naturally lead to a decrease in the wave heights in regions II and III. In the northern part of the Caspian Sea (region I), the change in the wind direction from southerly winds to westerly winds (Fig. 2) leads to an increase in the maximum fetch length and thus the wave heights are expected to increase in the region, which is confirmed in this study.

However, the observed wave height changes in the southern basin (regions IV and V) are not consistent with the observation that the predominant southerly winds were replaced by southwesterly ones in these regions (Fig. 2). This shift would have slightly increased the maximum fetch length from ~350 km to ~400 km and thus would lead to a certain increase in the wave heights in region V. However, satellite altimetry reveals a significant decrease in the wave heights in this region. This discrepancy calls for additional research to understand the reasons and patterns of changes in winds and waves in this area.

The pattern of changes in the extreme wave heights (meant here as 99th percentiles of wave heights) does not fully match the similar pattern for average wave heights. In particular, the average wave height has increased, but the extreme wave height has decreased in the eastern segment of the southern basin. Such seemingly inconsistent patterns can be produced by changing wind directions over elongated water bodies (Soomere et al. 2010). Therefore, a likely reason for the complicated pattern of spatial changes in the average and extreme wave heights may be a possible change in the directional structure of winds over different basins of the Caspian Sea.

**Acknowledgements.** The research was co-supported by the institutional financing by the Estonian Ministry of Education and Research (grant IUT33-3), the Flag-ERA project FuturICT2.0 and project AP05132939 'Control system design of the satellite formation motion for remote sensing of the Earth' of the al-Farabi Kazakh National University. We also acknowledge the support of the Horizon2020 Erasmus+ project CUPAGIS in terms of institutional collaboration and competence transfer. The visits of K. Kussembayeva to the Wave Engineering Laboratory were supported by the EU structural funds via the DoRa+ program. We thank ESIMO (World Ocean Status Government Information System, www.esimo.ru) for providing the *in situ* wave data for the Caspian Sea. We are grateful to L. Tuomi, L. Kelpšaitė-Rimkienė and the anonymous referee for constructive reviews of the manuscript. The publication costs of this article were partially covered by the Estonian Academy of Sciences.

# Kaspia mere lainekliima ajalis-ruumilised muutused aastail 2002–2013 satelliitaltimeetria alusel

Nadezhda Kudryavtseva, Kuanysh Kussembayeva, Zaure B. Rakisheva ja Tarmo Soomere

Kaspia mere lainekliima ruumiline muster ja selle ajalised muutused mere eri osades aastail 2002–2013 on kirjeldatud satelliitaltimeetria vahenditega. Aluseks on satelliidilt JASON-1 määratletud oluline lainekõrgus. Kaugseire vahenditega leitud lainekõrguste hinnangute tõenäosusjaotuse (Weibulli jaotuse) parameetrid paiknevad mere lõunaosas mõõdetud lainekõrguste samade parameetrite 95%-liste usaldusvahemike servas. Satelliidilt määratletud kuu keskmised lainekõrgused on märksa suuremad kui samas piirkonnas rannalt visuaalselt hinnatud lainekõrgused. Vahe pärineb osalt sellest, et satelliit ei detekteeri korralikult lainekõrgust suhteliselt rahulikul merel, ja osalt sellest, et ranna lähistel on lainekõrgus üldiselt väiksem kui satelliidi poolt vaadeldavas piirkonnas rannast eemal. On tuletatud meetod selle süstemaatilise erinevuse elimineerimiseks. Korrigeeritud satelliidiandmestiku ja kolmes kohas visuaalselt hinnatud kuu keskmiste lainekõrguste ruutkeskmise erinevus on 6 sentimeetrit.

Satelliitaltimeetria abil määratletud keskmine lainekõrgus on 0,5–0,7 m Kaspia mere põhjaosas, ligikaudu 1,2 m mere kesk- ja lõunaosas, kuid ulatub 1,8 meetrini mere keskosa lääneranniku lähistel Derbenti piirkonnas. Kogu mere keskmine lainekõrgus varieerus neil aastail 1,02–1,14 m vahel ilma selge trendita. Need arvud peegeldavad vaid suhteliselt intensiivset lainetust ja ülehindavad pikaajalist lainekõrgust ligikaudu 30% võrra. Statistiliselt olulised lainekõrguse muutused ilmnesid mere üksikutes osades. Lainekõrgus vähenes 0,019 ± 0,007 m/a mere keskosa idapoolses sektoris ja 0,04 ± 0,04 m/a mere lõunaosa läänepoolses sektoris. Need muutused on seletatavad muutustega valdavates tuulte suundades: läänetuulte sagedus on kasvanud ja lõunatuulte sagedus langenud. Ekstreemsed lainekõrgused (99%-iilid) nii kogu meres kui ka selle üksikutes osades varieerusid eri aastatel ulatuslikult, kuid ilma selge trendita. Mere lõunaosa idapoolses sektoris keskmine lainekõrgus kasvas, kuid ekstreemsed lainekõrgused kahanesid. Ka sellist laadi muutus on seletatav tugevate tuulte suuna muutustega.